\documentstyle[10pt, aas2pp4]{article}
\slugcomment{Feb. 1999, Accepted for publication in ApJ}
\lefthead{KIM, PARK \& LEE}
\righthead{STREAM-STREAM COLLISION AFTER TIDAL DISRUPTION}
\def\spose#1{\hbox to 0pt{#1\hss}}
\newcommand\lsim{\mathrel{\spose{\lower 3.0pt\hbox{$\mathchar"218$}}
     \raise 2.0pt\hbox{$\mathchar"13C$}}}
\newcommand\gsim{\mathrel{\spose{\lower 3.0pt\hbox{$\mathchar"218$}}
     \raise 2.0pt\hbox{$\mathchar"13E$}}}
\newcommand\msun{{\,M_\odot}}
\newcommand\rsun{{\,R_\odot}}
\begin{document}
\title{THE STREAM-STREAM COLLISION AFTER THE TIDAL DISRUPTION OF\\
A STAR AROUND A MASSIVE BLACK HOLE}
\author{Sungsoo S. Kim\altaffilmark{1}}
\affil{Department of Physics, Korea Advanced Institute of
Science \& Technology, Daejon 305-701, Korea}
\authoremail{sskim@space.kaist.ac.kr}
\author{Myeong-Gu Park}
\affil{Department of Astronomy and Atmospheric Sciences, Kyungpook National
University, Taegu 702-701, Korea}
\authoremail{mgp@kyungpook.ac.kr}
\and
\author{Hyung Mok Lee\altaffilmark{2}}
\affil{Department of Earth Sciences, Pusan National University, Pusan 609-735,
Korea}
\authoremail{hmlee@uju.es.pusan.ac.kr}

\altaffiltext{1}{This work has been initiated when he was at
Institute for Basic Sciences, Pusan National University, Korea}

\altaffiltext{2}{Current address: Department of Astronomy,
Seoul National University, Seoul 151-742, Korea; hmlee@astro.snu.ac.kr}

%%%%%%%%%%%%%%%%%%%%%%%%%%%%%%%%%%%%%%%%%%%%%%%%%%%%%%%%%%%%%%%%%%%%%%%%%%%%%%%%
\begin{abstract}
A star can be tidally disrupted around a massive black hole.  It has been
known that the debris forms a precessing stream, which may collide with itself.
The stream collision is a key process determining the subsequent evolution
of the stellar debris:  if the orbital energy is efficiently dissipated,
the debris will eventually form a circular disk (or torus).  In this paper,
we have numerically studied such stream collision resulting from the encounter
between a $10^6 \, \msun$ black hole and a $1 \, \msun$ normal star with
a pericenter radius of 100~$\rsun$.
A simple treatment for radiative cooling has been adopted for both
optically thick and thin regions.  We have found that approximately
10 to 15\% of the initial kinetic energy of the streams is converted
into thermal energy during the collision.
The spread in angular momentum of the incoming stream is increased
by a factor of 2 to 3,
and such increase, together with the decrease in kinetic energy,
significantly helps the circularization process.
Initial luminosity burst due to the collision
may reach as high as $10^{41} \, {\rm erg \, sec^{-1}}$ in $10^4$~sec, after
which the luminosity increases again (but slowly this time) to a steady
value of a few $10^{40} \, {\rm erg \, sec^{-1}}$ in a few times of $10^5$~sec.
The radiation from the system is expected to be close to Planckian with
effective temperature of $\sim 10^5$~K.
\end{abstract}


%%%%%%%%%%%%%%%%%%%%%%%%%%%%%%%%%%%%%%%%%%%%%%%%%%%%%%%%%%%%%%%%%%%%%%%%%%%%%%%%
\section{INTRODUCTION}
\label{sec:introduction}

There are dynamical evidences for supermassive black holes in nearby galaxies
including our own (e.g., \markcite{EG96}Eckart \& Genzel 1996).  Since these
black holes are embedded in a dense star cluster, tidal disruption of stars
can occur at a rate determined by the stellar density and velocity dispersion.
Rough estimates indicate around $10^{-4}$ per year for disruptive
encounters between a star and a massive black hole for the Milky way (e.g.,
\markcite{GL89}Goodman \& Lee 1989).

The debris from the disruptive encounter forms a precessing stream, and
self-interaction (collision between different parts of the stream)
may occur soon after the most tightly bound debris begins its second orbit.
\markcite{EC89}Evans \& Kochanek (1989),
\markcite{K93}Kochanek (1993), \markcite{L93}Laguna et al. (1993), and
\markcite{ML94}Monaghan \& Lee (1994) have tracked the debris from the
disruption to the collision phase.
The disruption can in principle lead to the flare up of luminosity
in galactic nuclei (e.g., \markcite{R88}Rees 1988) followed by an
accretion phase lasting for several years (\markcite{CLG1990}Cannizzo, Lee
\& Goodman 1990; \markcite{LU97}Loeb \& Ulmer 1997; \markcite{UPG98}Ulmer,
Paczy\'nski, \& Goodman 1998).  The studies on this rather long-lasting,
accreting phase started from the time when the debris is circularized assuming
that such circularization occurs in a fairly short amount of time (several
times of the orbital period of the most tightly bound debris).
The present study targets the evolution of the debris during and right after
the collision.  The circularization process will take much longer than
the time scale that we study here, and is beyond the scope of the present study.

The detailed evolution of the stellar debris has not been
studied carefully because of the intrinsic complexity of the problem.
The difficulty stems from the huge dynamic range of the physical parameters
of the debris around the black hole. For example, the pericentral distance
of the debris' orbit is typically 1 AU while the apocenter distance could be
of order of pc.  The most important phenomenon in determining the fate of
stellar debris is the collision between the incoming and outgoing streams
due to relativistic precession (e.g. \markcite{R88}Rees 1988,
\markcite{CLG1990}Cannizzo et al. 1990).  The supersonic collision between
two streams can in principle convert orbital energy into thermal energy.
If this process efficiently removes the orbital energy, the stream transforms
itself into a circular disk (or torus) around the black hole.
The stream crossing is likely to be intermittent in nature because
the collision can destroy the stream geometry (\markcite{LKR95}Lee, Kang,
\& Ryu 1995; LKR hereafter).  Such a cycle will repeat with period of
free-fall time at the crossing point.

LKR have numerically studied the stream
collisions and showed that the shock should be able to convert the ordered
orbital motion of two fluids into expansion of a bubble and thus produce
the torus around the black hole.  For simplicity, LKR assumed that the
collision is adiabatic and gas pressure is dominant over radiation pressure.
However, it is more likely that the radiation pressure is more important at
the shocked region and radiative cooling may play an important role.
The purpose of the present study is to extend
LKR's numerical simulations including radiative processes.  We pay special
attention to the physical status of shocked region and expanding post-collision
gas, the amounts of thermalization and radiative cooling, and changes in
orbital energy and angular momentum of the streams.

We describe our numerical methods for hydrodynamics and radiative cooling
in \S \ref{sec:methods}.  Simulation results and energy conversion issue are
discussed in \S \ref{sec:results} and some estimates for orbital and
radiative evolution are made in \S \ref{sec:discussion}.  The final section
summarizes our findings.

%%%%%%%%%%%%%%%%%%%%%%%%%%%%%%%%%%%%%%%%%%%%%%%%%%%%%%%%%%%%%%%%%%%%%%%%%%%%%%%%
\section{NUMERICAL METHODS}
\label{sec:methods}

\subsection{Hydrodynamic Simulation}
\label{sec:hydrodynamics}

We perform simulations of supersonic collisions between two gas
streams with a three-dimensional hydrodynamics code based on the total
variation diminishing (TVD) scheme (\markcite{H83}Harten 1983;
\markcite{R93}Ryu et al. 1993), which
is an explicit, second-order, Eulerian finite-differencing scheme.  The
code used in LKR has been modified here to include the effects
of radiative cooling.  The treatment for the radiative cooling is discussed
in the next subsection.

While hydrodynamic simulations in LKR were performed purely dimensionlessly,
the introduction of radiative processes requires some dimensional constraints.
In this paper, all simulation variables have specific physical values.
We targeted the stream debris resulting from the encounter between a
$10^6 \, \msun$ black hole and a normal star with mass of 1 $\msun$ and
radius of 1~$\rsun$, which initially has a parabolic orbit with a
pericenter radius of 100~$\rsun$.

Some initial conditions of our simulations were obtained from the results
by \markcite{K94}Kochanek (1994) and \markcite{LK96}Lee \& Kim (1996).
Velocities and mass fluxes of two streams were
obtained from the values at $t \approx 1750 \, t_D$ of his Figure 6 ($v_1
= 0.0147 \, {\rm c}$, $v_2 = 0.0083 \, {\rm c}$, $\dot M_1 = 1 \msun
{\rm yr^{-1}}$, and $\dot M_2 = 0.5 \msun {\rm yr^{-1}}$).  We have
chosen this epoch because $\dot M_2$ becomes significant compared to
$\dot M_1$ ($\dot M_2 \approx \dot M_1/2$) at this epoch (subscript 1 is
for the stream making its first orbit, stream 1, and 2 is for the stream
making its second orbit, stream 2).  Kochanek found that the angle between
the streams at the collision point is typically $130 \arcdeg$ to $140 \arcdeg$
and here we adopted a collision angle of $135 \arcdeg$.  
However, his results
for the widths of two streams $\Delta_1$ and $\Delta_2$
at a certain epoch depend on the way of treating the viscosity.
For this reason, we perform simulations for five different sets of $\Delta$'s.
We have fixed the heights of two streams, $H_1$ and $H_2$ except
for one run because these quantities are less dependent on the viscosity
treatment.

Model parameters of our simulations are shown in Table \ref{table:stream}.
Our standard model is run 7, where the cross sections of the streams are
round ($\Delta_1=H_1$ \& $\Delta_2=H_2$), two streams move in the same
$z=50$ plane (hereafter the coordinate positions are represented by the pixel
number; see $l_{cell}$ below for the physical size of each pixel),
and the mass flux and the radius of stream 1 are 2 and 5/3 times larger
than stream 2, respectively.  Runs 2, 8, 9, and 10 are for different
$\Delta$ and $H$ values.  Runs 11 and 12, where stream 2 is set to move
in $z=54$ and 58, respectively, are for the estimation of the effects of
off-centered collision that could occur when there is Lense-Thirring
precession.  Finally, to see the effects of radiation, we will compare
runs 7 and 1, where radiative cooling is not considered.

The computational box of our simulations consists of $101^3$ cubical cells.
The physical length of each cell $l_{cell}$ is $5.93 \times 10^{10} \, \rm{cm}$.
The gas is assumed to be ionized already before the collision and
the initial pressure of the streams is set such that the Mach number of the
stream is 300.  This gives an initial temperature of about
$2 \times 10^4 \, {\rm K}$ for stream 1 and $7 \times 10^3 \, {\rm K}$ for
stream 2.  The initial density of the ambient medium
is $10^{-4}$ times smaller than the average density of the streams
and the initial temperature of the ambient medium is set to be slightly
smaller than that of the streams.  The initial density profile
of stream's cross section has a functional form of $\exp(-4y^2/\Delta^2 -
4z^2/H^2)$.  Unlike the simulations of LKR, where two streams are set to
collide on their front ends, our stream 2 is set to collide on the side
of stream 1, which more resembles the geometry of actual stream collisions.
Streams 1 and 2 are set to keep flowing into the simulation box until
the end of the simulation.

The total integration time of our simulations is $4 \times 10^4$~sec.
\markcite{K94}Kochanek (1994) finds that for the encounter with parameters
adopted in this paper, the debris stream will self-intersect at
5000-6000~$\rsun$ from the black hole on its second orbit and the first
collision phase will last for $\sim 10^6 \, {\rm sec}$.  Thus the
one-dimensional size of our simulation box is about 100 times smaller
than the distance to the black hole from the collision region and our
simulations cover only small fraction of the first collision phase.

\subsection{Radiative Processes}
\label{sec:radiation}

The average particle number density of the streams is of order $10^{17} \,
{\rm cm^{-3}}$ and the characteristic radius of the streams is of order
$10^{12} \, {\rm cm}$.  For the typical column density of $\sim 10^{29} \,
{\rm cm^{-2}}$, the electron scattering optical depth $\tau_{es}$ is much
larger than unity, and so is the effective optical depths $\tau_* \equiv
[3 \tau_{ff} (\tau_{ff} +\tau_{es})]^{1/2}$ (\markcite{RL79}Rybicki \&
Lightman, 1979), where $\tau_{ff}$ is the
Rosseland mean free-free absortion optical depth.  Thus the post-collision
region may be approximated by thermal equilibrium where matter and
radiation are strongly coupled.  The total internal energy density, $u$, is now
the sum of the gas thermal energy density, $u_g$, and the radiation energy
density, $u_r$:
\begin{equation}
\label{totalu}
      u = u_g + u_r = {nkT \over \gamma_g -1} + a T^4,
\end{equation}
where $n$ is the gas number density, $k$ the Boltzmann constant, $a$ the
radiation constant, and $\gamma_g$ (=5/3) the adiabatic index for gas.
The mixture of gas and radiation gives an effective adiabatic index
$\gamma_{eff}$ between 5/3 and 4/3, depending on the ratio $u_g / u_r$
(or on the ratio $P_g / P_r$, where $P_g$ and $P_r$ are the pressure
of the gas and the radiation respectively).

\begin{figure}[p]
%Fig 1
\centerline{\epsfxsize=8.8cm\epsfbox{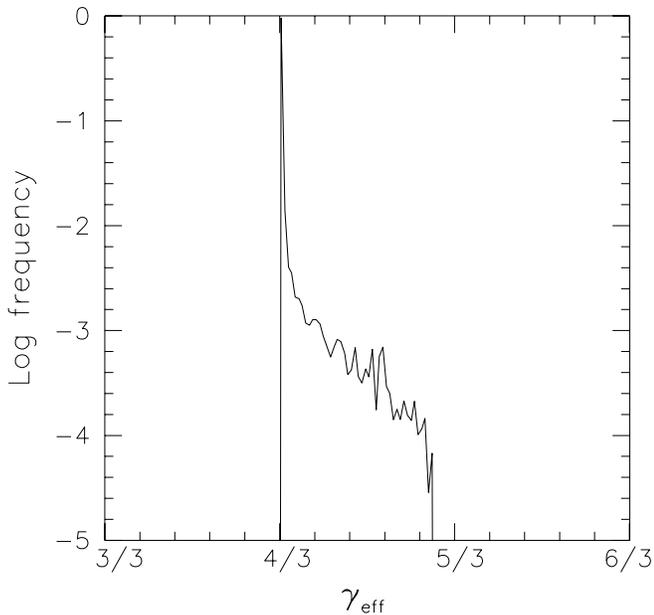}}
\caption
{\label{fig:gamma}Normalized, density-weighted frequency distribution of
post-collision $\gamma_{eff}$ for run 7 at $t = 4 \times 10^4$~sec.
Most PCG has $\gamma_{eff}$ very close to 4/3,
the adiabatic index value for radiation.}
\end{figure}

We assume that the post-collision region has physical conditions $P_r \gg P_g$
and adopt $\gamma_{eff}=4/3$ except for run 1.  The validity of this assumption
may be checked in Figure \ref{fig:gamma}, which illustrates the density
weighted distribution of post-collision $\gamma_{eff}$ for run 7 at $t=4 \times
10^4$~sec.  Here, $\gamma_{eff}$ is defined by
\begin{equation}
\label{newT}
	{P_{tot} \over \gamma_{eff}-1} \equiv {P_g \over \gamma_g -1} +
						{P_r \over 4/3 -1},
\end{equation}
where $P_g=nkT$ and $P_r={1 \over 3}aT^4$.  Most of the thermalized material
falls onto the region where $\gamma_{eff}$ is close to $4/3$.

Since the generic temperature of the post-collision
region is well above $10^4$~K, we assume that gas is completely ionized and is
cooled by bremsstrahlung when $\tau_* \lsim 1$.  Between $10^4$~K and $10^5$~K,
line cooling by various atoms could be significant compared to bremsstrahlung.
However, the opacity for line photons is usually much larger than that for
the continuum photons and therefore do not contribute much to actual cooling.

Bremsstrahlung photons can not escape freely either in most cases.
They have to diffuse out from the inner, generally hotter, region to
the outer, generally cooler, region, and in doing so heat up the outer parts.
As in the stellar interior, the amount of diffusion
depends on the temperature gradient,
and the temperature gradient itself is maintained by the amount
of energy transported from one part of the gas to the other.
However, one important difference between colliding gas streams
and the stellar interior is that part of the bulk kinetic energy
can always be converted to the thermal energy in any part of the
flow which is ultimately radiated away, whereas
only the nuclear burning core generates the energy and
outer envelope passively transfers the radiation in the stellar interior.
Hence, the correct way to handle this kind of dynamic mixture of gas
plus radiation is to fully solve the three-dimensional
radiative transfer at a given instant along with hydrodynamics.
This will tell us how much radiation is generated and transported from
one part to the other and, therefore, determine the physical state of the gas
and radiation in the next time step. However, this is quite a formidable task
and we need to find a simpler way to treat this.

One reasonable and efficient way for this type of hydrodynamic simulation
is to use a volume cooling rate which can approximate the radiation transport
process.  We adopt the following form for the cooling (\markcite{LW91}Liang
\& Wandel 1991; \markcite{WL91}Wandel \& Liang 1991):
\begin{eqnarray}
\label{emit1}
      \epsilon = \eta \beta \, {T^4 \over R_*} \, { 1 - e^{-\tau_*} \over
                 e^{-\tau_*} + (1-e^{-\tau_*}) {\eta \tau_* /(12 \sigma)}},
\end{eqnarray}
where
\begin{eqnarray}
      \eta \equiv {\epsilon_{ff} \over \alpha_{ff} T^4 }
                       = 8.2 \times 10^{-3} {{\bar g_B} \over {\bar g_R}}; \; \;
      \beta \equiv \sqrt{\alpha_{ff} \over 3(\alpha_{ff}+
                       \alpha_{es})}.
\end{eqnarray}
Here $\epsilon$ is the cooling rate per unit volume,
$\sigma$ the Stefan-Boltzmann constant, $\bar g_B$ the frequency
average of the velocity averaged free-free Gaunt factor, and $\bar g_R$
the Rosseland mean free-free Gaunt factor.  Also, $\alpha_{ff}$ and
$\alpha_{es}$ are the Rosseland mean free-free absorption and
electron scattering coefficients, respectively.  This form is valid both
for effectively optically thin ($\tau_* \ll 1$) and for effectively
optically thick ($\tau_* \gg 1$) cases.

For effectively optically thin cases, equation (\ref{emit1})
becomes the usual brensstrahlung emission rate per unit volume,
\begin{eqnarray}
\label{emit3}
	\epsilon = \epsilon_{ff} & & {\rm for} \; \; \tau_* \ll 1.
\end{eqnarray}
For effectively optically thick cases, equation (\ref{emit1})
reduces to the diffusive cooling rate by thermalized gas,
\begin{eqnarray}
\label{emit2}
      \epsilon = { u_r(T) \over \tau_{tot} R_{tot} /c}
                 & & {\rm for} \; \; \tau_* \gg 1,
\end{eqnarray}
where the total optical depth $\tau_{tot} \equiv \tau_{ff} + \tau_{es}$
and $R_{tot} \equiv \tau_{tot} / (\alpha_{ff} +\alpha_{es})$.
Integration of $\epsilon$ over the whole volume yields roughly the surface
area times $\sigma T_{eff}^4$, where $T_{eff}^4 \simeq T^4 \tau_{tot}$.
This is equivalent to finding the effective temperature of the
photosphere using the Eddington approximation, and summing up the flux
over the whole surface.

The amount of cooling is calculated explicitly by
equation (\ref{emit1}), and is subtracted from the total internal energy
of each cell after every hydrodynamic step.  To obtain optical depths,
appropriate absorption or scattering coefficients were integrated along
$\pm x$, $\pm z$ directions, and the smallest among the integrated
values is chosen.  The effective photon travel length $R_*$ is determined
by $\tau_* / [3 \alpha_{ff} (\alpha_{ff} +\alpha_{es})]^{1/2}$, where
local values were used for the calculation of $\alpha_{ff}$ and $\alpha_{es}$.

The exact time-dependence of escaping photons from optically thick medium
was solved by \markcite{ST80}Sunyaev \& Titarchuk (1980) for a spherical
plasma cloud.
Their work shows that when a point source, located at the center
of the cloud of radius $R$ and the optical depth $\tau$, emits
flash of radiation at time $t=0$, the photon escape rate is highest
when $t \simeq t_{esc}$ and exponentially declines
afterward with the same timescale $t_{esc}$, where $t_{esc} = (3/\pi^2) t_d$
and $t_d \equiv (R/c)\tau$ is the photon diffusion timescale. However,
when the sources of photons are homogeneously distributed within the 
sphere, photon escape rate is highest at $t=0$ and declines afterward
with the same exponential behaviour.
When the distribution of photon sources follows a sine curve, the exact
solution of the escape probability is found to be $\exp(-t/t_{esc})$.
Therefore, the fraction of photons that escaped from $t=0$ to
$\Delta t$ ($\ll t_{esc}$) is $(\pi^2/3)[\Delta t/(\tau R/c)]$.
This is equivalent to equation (\ref{emit2}) with a
different constant factor. Since sources of photons in our
simulation are distributed but do not follow the exact sine
form, we will set this constant factor as unity in our simulations.

Energy of emitted photons can be changed via Compton scattering
since electron scattering optical depth is usually quite high.
However, this happens only when the typical length scale for absorption
is longer than that for Compton upscatter. Otherwise, photons will be
absorbed and thermalized before being significantly upscattered.
Since the mean free path due to
electron scattering is $(n_e \sigma_{es})^{-1}$
and photons need to be scattered several $4 m_e c^2 / k T $
times to gain significant energy boost, the typical length scale for
upscattering is
$ \lambda_{\rm Compt} \equiv (n_e \sigma_{es} 4 k T / m_e c^2)^{-1}$.
Here, $n_e$ is the electron number density,
$\sigma_{\rm es}$ the Thomson cross section, and $m_e$ the mass
of electron. In the thermalized regions of our simulations, the absorption
length scale $\lambda_{\rm abs} = \alpha_{ff}^{-1}$ is much larger than
$\lambda_{\rm Compt}$ and Compton scattering can be neglected.

It is possible in high temperature plasma that ions decouple from electrons
and each develops widely different temperature.  If so, the gas would have to
be treated as such.  During the collision, ions and electrons may reach the
highest possible temperature, which is close to the virial temperature:
for ions, $T_i^{max} \equiv m_p v^2/3k$, and for electrons, $T_e^{max}
\equiv m_e v^2/3k$, where $m_p$ and $m_e$ are the proton and electron
mass, respectively, and $v$ the collision velocity.  With our initial
stream parameters, we find that the Coulomb time scale between ions and
electrons at virial temperature, $\sim 10^{-7}$~sec (Park 1990), is a few
orders of magnitude smaller than the bremsstrahlung cooling time scale,
$\sim 10^{-3}$~sec.  Besides, the collision region is effectively optically
thick at this electron temperature, and ions will be stronly coupled with
electrons and radiation will be locked with electrons at all times.

%%%%%%%%%%%%%%%%%%%%%%%%%%%%%%%%%%%%%%%%%%%%%%%%%%%%%%%%%%%%%%%%%%%%%%%%%%%%%%%%
\section{SIMULATION RESULTS}
\label{sec:results}

\subsection{Morphology and Structure}
\label{sec:morphology}

\begin{figure}[p]
%Fig 2
\centerline{\epsfxsize=8.8cm\epsfbox{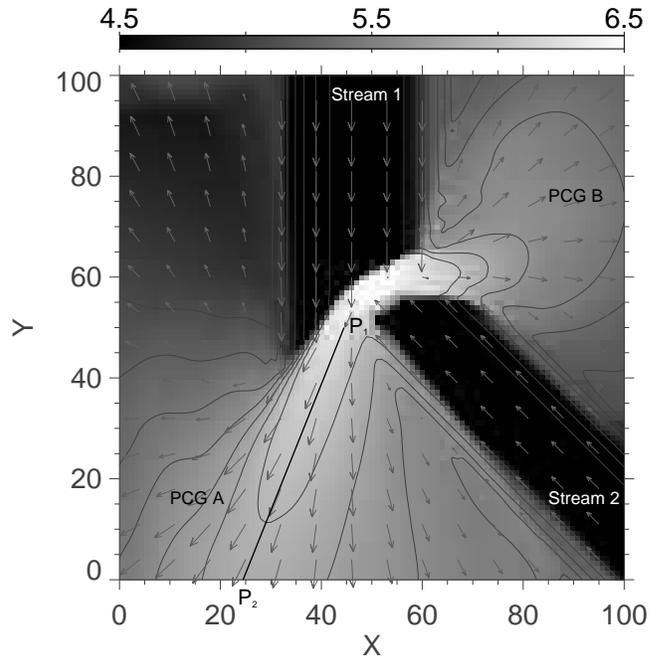}}
\caption
{\label{fig:slice7}Density, temperature, velocity vector plot of $z=50$ plane
for run 7 at $t = 4 \times 10^4$~sec. Units of $x$, $y$, and $z$ are the grid
size of the simulation, $5.93 \times 10^{10}$~cm.  The contour is for density,
grey-scale map for temperature, and arrows for velocity vectors projected
onto the x-y plane.  The temperature scale is shown as a bar at the top
of the Figure.  Stream 1 moves from top to bottom at $x=50$ and stream 2
moves from lower-right corner to upper-left corner.  Two streams are
colliding at the center of the Figure, and thermalized PCG
is expanding out mainly toward lower-left corner.  Profiles of the line
connecting points P$_1$ and P$_2$ are shown in Figure \ref{fig:ycut7}.}
\end{figure}

Figure \ref{fig:slice7} shows $n$ (contour), $T$ (greyscale), and $\vec v_{xy}$
(velocity projected on the x-y plane; arrow) in $z$=50 plane of run 7 at
$t=4 \times 10^4$~sec.  Stream 1 coming from the top boundary is colliding near
$x,y=(55,60)$ with stream 2 coming from the lower-right corner.  Mass
elements are almost instantly thermalized in the collision area and are
spread out in two directions.  While part of stream 2 is bounced by stream 1
and heads for $+x$, $+y$ direction, the rest merges with stream 1,
which is deflected toward $-x$ direction after the collision.

The thermalized gas emerges from the narrow, slab-like shock region,
and expands out into a larger volume in two opposite directions due to
the post-collision pressure.  Thus the post-collision material forms two
expanding streams several times thicker than the pre-collision streams.
Hereafter, the post-collision gas (PCG) in the -y direction from the collision
region will be called PCG A, and the other one in the opposite direction
will be called PCG B.

\begin{figure*}[p]
%Fig 3
\centerline{\epsfxsize=15cm\epsfbox{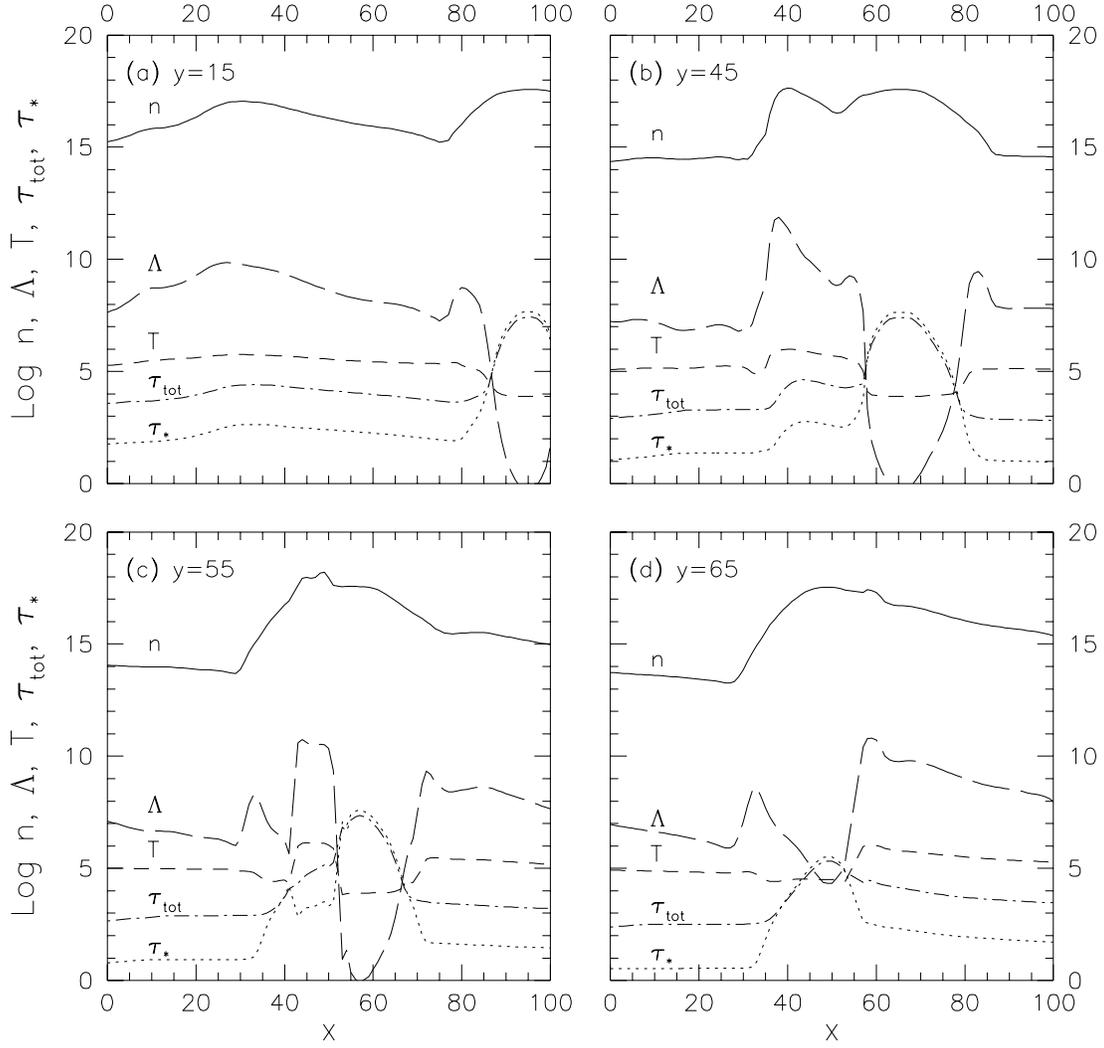}}
\caption
{\label{fig:cut7}Number density ($n$), radiative cooling per cell ($\Lambda$),
temperature ($T$), total optical depth ($\tau_{tot}$), and effective optical
depth ($\tau_*$) profiles of z=50, y=15 (a), y=45 (b), y=55 (c), and y=60
(d) lines for run 7 at $t = 4 \times 10^4$~sec.  The units for $n$, $\Lambda$,
and $T$ are ${\rm cm^{-3}}$, $10^{25} {\rm erg \, /sec}$, and K, respectively.}
\end{figure*}

Profiles of $n$, $T$ (defined as eq. [\ref{totalu}]), radiative
cooling per cell ($\Lambda$), $\tau_{tot}$, and $\tau_*$ in $z$=50 plane of
run 7 at $t=4 \times 10^4$~sec are plotted in Figure \ref{fig:cut7} for four
different values of $y$.  Since we did not consider the cooling under
$T=10^4$~K, $T$ is not allowed to become below $10^4 \,{\rm K}$.
Also, $T$ scarecely goes higher than $10^6 \,{\rm K}$ because cooling becomes
more efficient at higher temperature.  On the other hand, $\tau_*$ is greater
than 10 in most regions because of the high Thomson opacity, and goes
up as high as $\sim 10^7$ in the regions that are not thermalized yet.

Figure \ref{fig:cut7}(c) shows the profile of $y$=55, $z$=50 line which
includes the central region of the collision.  The shock front is formed at
$40 < x < 50$, where temperature is one to two orders of magnitude higher
than that of pre-collision gas.  Cooling is most prominent at the shock slab,
but the ratio of the cooling to the total internal energy (gas + radiation)
in that region during one simulation time step is only $\sim~0.01~\%$.
Therefore the radiative cooling time scale is considerably longer than
the hydrodynamical time scale (one simulation time step).

\markcite{K94}Kochanek (1994) found that the temperature of
the streams right before the collision is well below $10^4 \, \rm{K}$.
Since we do not consider cooling processes under $10^4 \, \rm{K}$ in our
simulations, no cooling is expected in the pre-collision gas.  However,
pre-collision stream regions, such as $55 < x < 80$ of (b) and $x > 80$
of panel (a) of Figure \ref{fig:cut7}, show some cooling
although very small.  This is because the initial temperature of the streams
is $\sim \, 2 \times 10^4 \, \rm{K}$, which is determined by setting
Mach number equal to 300.  The stream material of this temperature will
already begin to experience radiative cooling even before the collision,
which is not desirable in our simulations.  Higher Mach number would give
lower initial stream temperature, but numerical difficulties prohibit us to use
such high Mach numbers.  In any case, the amount of such undesirable cooling is
negligible compared to the one in the pre-collision gas.

At the boundaries between the expanding PCGs and cold streams, such as
at $x \approx 80$ of panel (a), $x \approx 85$ of panel (b), and $x \approx 30$
of panel (d), the flows of the PCG and the stream move in opposite directions
and shear viscosity thermalizes the boundary region between the flows.  Since
such regions have relatively small optical depths, thermalized energy is
efficiently radiated away and local cooling maxima occur in those regions.

\begin{figure}[p]
%Fig 4
\centerline{\epsfxsize=8.8cm\epsfbox{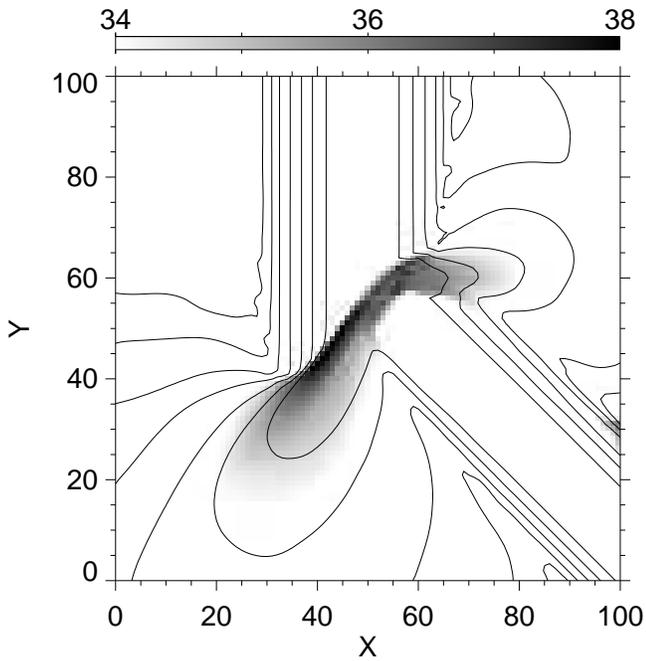}}
\caption
{\label{fig:cslice7}Cooling map in grey-scale for $z=50$ plane of run 7
at $t = 4 \times 10^4$~sec.  Overplotted is the density contour, and
a grey-scale bar is shown at the top in units of $\rm{erg \,/sec}$.}
\end{figure}

A grey scale map of cooling in $z$=50 plane for the same simulation at the
same epoch as Figure \ref{fig:slice7} is plotted in Figure \ref{fig:cslice7}
along with the density contour map.  Cooling is most efficient at the collision
region and at the mouths of two PCGs.
Although our simulations do not allow the exchange of photons between
gas elements, we may anticipate that the photons created in the collision
region will preferentially escape into the PCGs rather than into the incoming
cold streams because the expanding PCGs have much lower density and
thus play the role of window to the escaping photons.  The leakage of the
thermal energy in the collision region into the cold, incoming streams will be
inefficient because the cold streams have huge optical depths.

\begin{figure}[p]
%Fig 5
\centerline{\epsfxsize=13cm\epsfbox{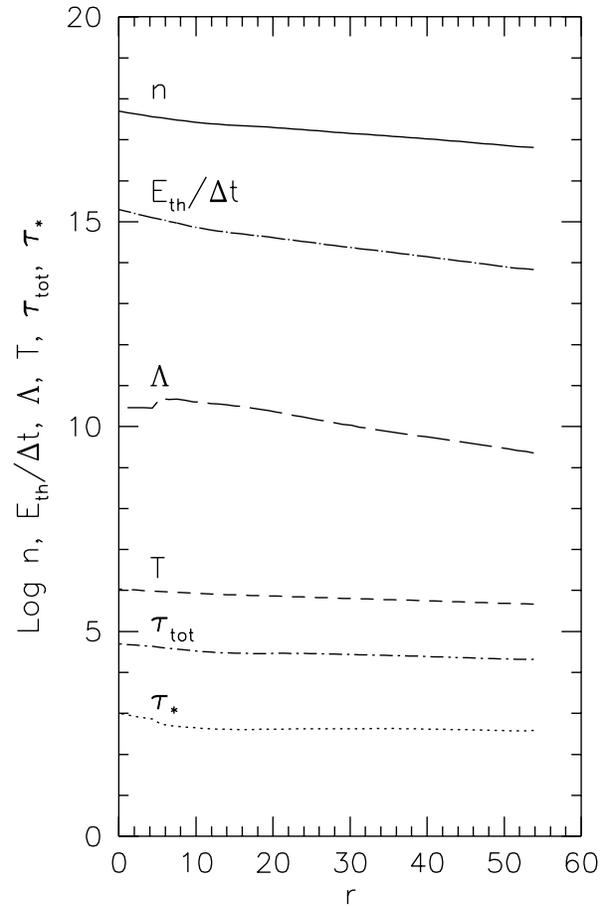}}
\caption
{\label{fig:ycut7}Number density ($n$), total internal energy per cell
divided by a time step ($E_{th}/\Delta t$), cooling per cell ($\Lambda$),
temperature ($T$), total optical depth ($\tau_{tot}$), and effective optical
depth ($\tau_*$) profiles along the line connecting points P$_1$ and P$_2$
in Figure \ref{fig:slice7}.  The units are same as in Figure \ref{fig:cut7}
with $E_{th}/\Delta t$ having the same units as $\Lambda$.}
\end{figure}

Profiles along the line connecting points P$_1$ and P$_2$ in $z$=50 plane of
Figure \ref{fig:slice7} are shown in Figure \ref{fig:ycut7}, where $r$ is the
distance in $l_{cell}$ from point P$_1$.  Note that the density profile shows
an exponential behavior at $r>20$.  Since $T$ and $\Lambda$ are tightly related
to $n$, these variables also decrease exponentially in the same region.
The profiles of these variables are found to be well fitted by $n \propto
\exp(-r/1.7 \times 10^{12} \, {\rm cm})$, $T \propto \exp(-r/4.9
\times 10^{12} \, {\rm cm})$, and $\Lambda \propto \exp(-r/9.2 \times 10^{11}
\, {\rm cm})$.

As we will discuss in the following subsection, the fraction of the energy
radiated in the simulation box to the initial kinetic energy is only
$\sim 0.3$~\%.
Thus the dependence of $T$ on $n$ will not be far from the adiabatic relation,
$T \propto n^{\gamma -1} \propto n^{1/3}$ (recall that we adopted $\gamma
= 4/3$ for all our simulations).  Indeed, the $T$ scale length is almost one
third of the $n$ scale length.  If the PCG is assumed to expand
in two dimensions (as when a stream with infinite length expands to the width
and height directions), one would have $R_{tot} \propto n^{-1/2}$.
Then the $\Lambda$ scale length, $9.2 \times 10^{11}
\, {\rm cm}$, is a consequence of the cooling formula for optically thick
region, equation (\ref{emit2}): $\Lambda \propto T^4 R_{tot}^{-1}
\tau_{tot}^{-1}$$\propto T^4 R_{tot}^{-2} \alpha_{es}^{-1/2}$$\propto
T^4 R_{tot}^{-2} n^{-1/2}$ $\propto \exp(-r/9.0 \times 10^{11} \, {\rm cm})$,
where $\tau_{es} \gg \tau_{ff}$ has been assumed.

\begin{figure}[p]
\vspace{3cm}
%Fig 6
\centerline{\epsfxsize=8.8cm\epsfbox{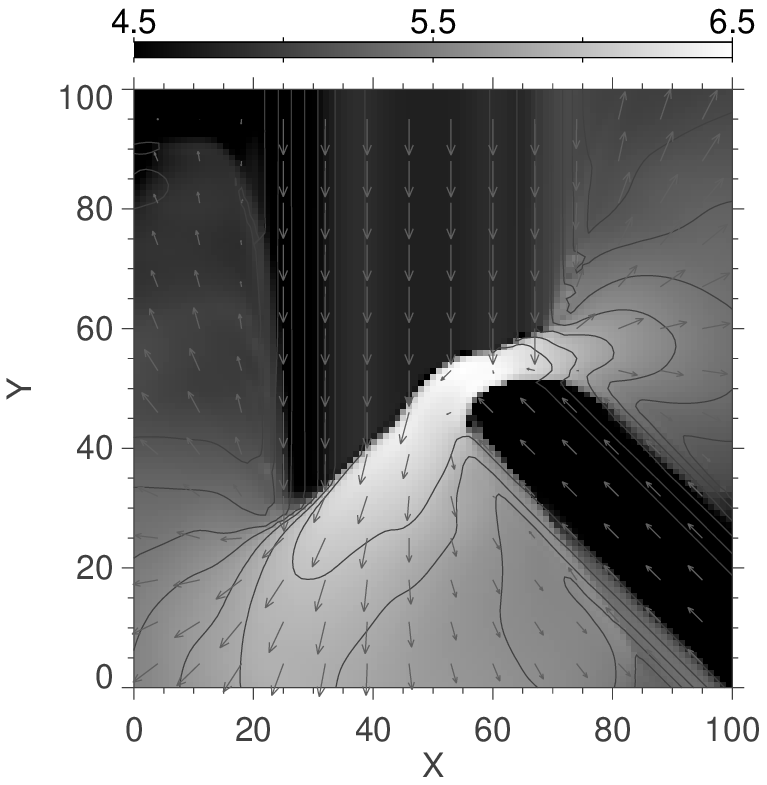}}
\caption
{\label{fig:slice8}Density, temperature, velocity vector plot of $z=50$ plane
for run 8 at $t = 4 \times 10^4$~sec. See Figure \ref{fig:slice7} for
explanations.}
\vspace{3cm}
\end{figure}

\begin{figure}[p]
%Fig 7
\centerline{\epsfxsize=8.8cm\epsfbox{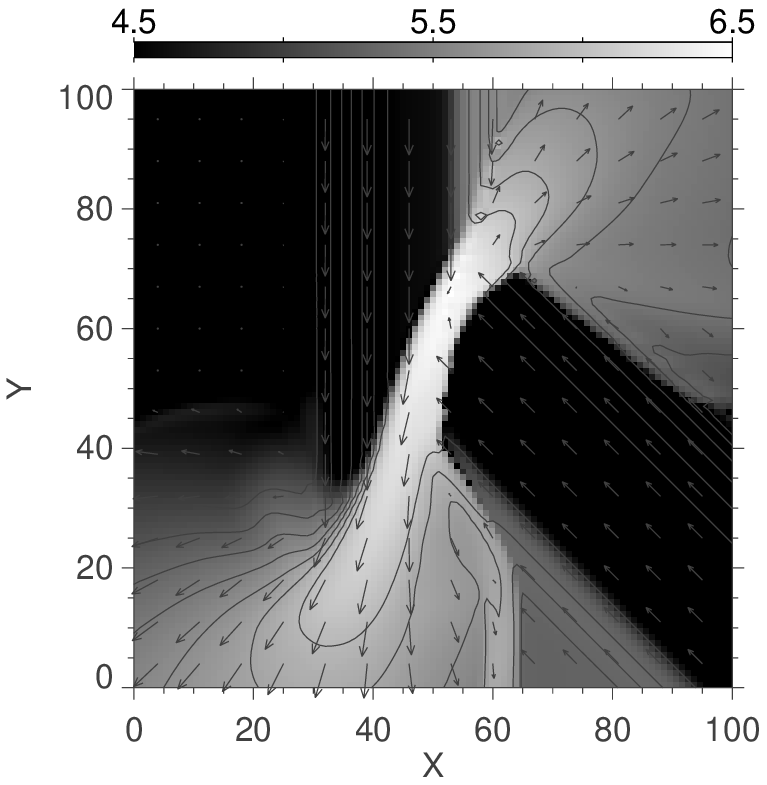}}
\caption
{\label{fig:slice9}Density, temperature, velocity vector plot of $z=50$ plane
for run 9 at $t = 4 \times 10^4$~sec. See Figure \ref{fig:slice7} for
explanations.}
\end{figure}

Simulations representing two opposite $\Delta_1 / \Delta_2$ cases, runs 8 and
9, show very distinct morphological evolution as shown in Figures
\ref{fig:slice8} and \ref{fig:slice9}.
First, while the structure of the thermalized slab in the collision region
of run 8 is considerably bent toward stream 2, that of
run 9 is nearly straight with only a small curvature toward stream 1.
Such a difference is a consequence of different mass flux per unit area,
$\dot M / dA$.  In run 8, stream 2 has a greater $\dot M/dA$ than that of
stream 1, thus it penetrates into stream 1 more deeply than in run 9
and forms a shock front almost perpendicular to both streams although
the greater $\dot M$ of stream 1 determines the eventual direction of
PCG.  In run 9, on the other hand, $\dot M/dA$ of stream 2 is
significantly smaller than that of stream 1, therefore stream 2 is effectively
bounced at the long shock front.
The expansion aspects of their PCGs are accordingly different.
The PCG A of run 8 is largely extended, but that of
run 9 maintains its slab-like shape until it escapes from the simulation box.
Thus the orbital evolution of the PCG A is expected to be considerably
different for two simulations, and will be discussed again in
\S \ref{sec:discussion}.

\begin{figure}[p]
%Fig 8
\centerline{\epsfxsize=8.8cm\epsfbox{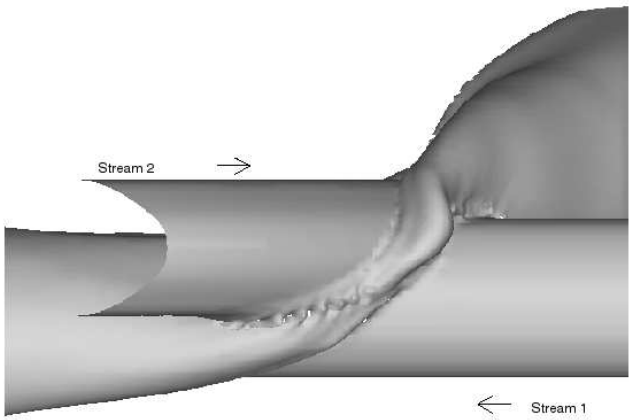}}
\caption
{\label{fig:iso12}Three-dimensional isodensity surface plot for run 12
at $t = 4 \times 10^4$~sec.  The surface represents $n = 8 \times 10^{15}
\rm{cm^{-3}}$ points.  The view angle is from $+x$ axis to $-x$ axis.  Stream 1
moves from right to left in $z=50$ plane, and stream 2 moves from left to
right with an angle of $45\arcdeg$ into the Figure in $z=58$ plane.}
\end{figure}

If the central black hole is rotating, the orbital plane of the stream
could experience the Lens-Thirring precession that causes slightly
off-centered collision.
Figure \ref{fig:iso12} shows an isodensity plot at $t=4 \times 10^4$~sec
for run 12, where only the lower part of stream 2 is set to collide
with the upper part of stream 1.  The collision produces an extended PCG A
and a narrow PCG B with a pitch angle of $\lsim 45 \arcdeg$ from x-y plane.

\subsection{Energy Conversion}
\label{sec:energy}

A certain fraction of stream's kinetic energy is converted into thermal
energy due to supersonic collision and the key interest of the present study
is how much fraction of such thermal energy escapes through radiative
cooling.  If the collision significantly alters the orbital motion of
stream 1, the collision phenomena becomes intermittent, and the first
collision would last for about $10^6 \, \rm{sec}$.  However, since the
hydrodynamic simulations in this study are limited within a small region
around the collision point, the duration of simulations with reasonable
accuracy is only order of the time scale for the stream to cross the
computational domain,
which is few percents of the duration of the first collision phase.
For this reason, we perform our simulations until energy conversion
enters a quasi steady state, and extrapolate the results to make some estimates
on the energy budget problem during one whole collision phase.

Since mass elements are continuously coming into and going out of the
calculation volume, and since it is very difficult to keep track of mass
elements in the simulations based on fixed data points (grids), calculating
the amount of energy conversion of a mass element which experiences
collision is not trivial.  Therefore, we estimate the energy budget in a
time-cumulative way: $E_{in,sum}$ is defined as the cumulated (from $t=0$)
input energy ($E_{in,sum}=0$ at $t=0$), $E_{th,sum}$ is the summation of
the total internal energy (gas+radiation) inside the calculation box at a
certain epoch and cumulated total internal energy which has escaped from the box
until that epoch, and $E_{r,sum}$ is the cumulative energy which has been
radiated away inside the simulation box until the same epoch.

\begin{figure}[p]
%Fig 9
\centerline{\epsfxsize=14cm\epsfbox{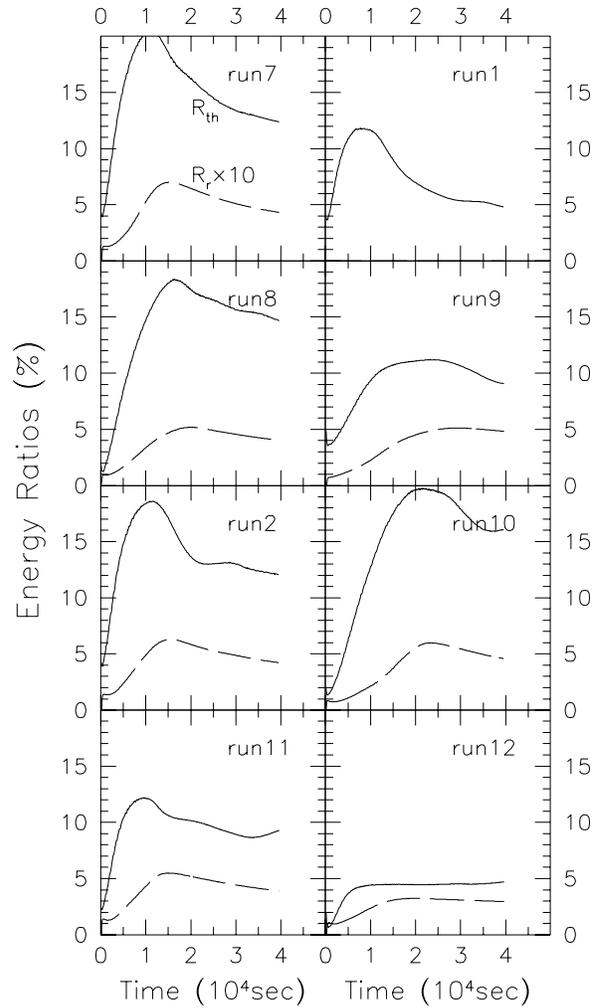}}
\caption
{\label{fig:energy8b}Evolution of the ratio of total thermal energy
to total input kinetic energy ($R_{th}$; solid lines) and the ratio of total
cooling inside the simulation box to total input kinetic energy ($R_{r}$;
dashed lines) of all simulations.}
\end{figure}

The total internal energy of a mass element before
collision is negligible compared to its kinetic energy.  Thus the ratio
$R_{th} \equiv (E_{th,sum}+E_{r,sum})/E_{in,sum}$ is the fraction of total
thermal energy converted from the total input kinetic energy.
And the ratio $R_{r} \equiv E_{r,sum}/E_{in,sum}$ is the fraction of
total radiated energy inside the simulation box out of the total input energy.
The evolution of $R_{th}$ and $R_{r}$ are shown in Figure \ref{fig:energy8b}.
After a rapid increase in the beginning, $R_{th}$
decreases to an asymptotic value slowly.
This relatively low thermalization rate in the later part is due to a growth
of accumulated mass elements in the collision area which enlarges the
effective collision cross section and lessens the relative velocity
between two streams.  In case of run 7, $R_{th}$ reaches as high as 20~\% in
the beginning, then it decreases to an asymptotic value of about 12~\%.
On the other hand, the cooling fraction $R_{r}$ of the
same simulation has a gentle maximum at $t \approx 1.5 \times 10^4 \,
{\rm sec}$ and slowly converges to $\sim$~0.4~\%.  From
these ratios, we expect that the overall amount of energy that
escapes from our calculation region as radiation during the whole
collision phase is $\sim$~0.4~\% of the total energy input or
$\sim$~3~\% of total thermal energy.  Note that $R_r$ is just the `inside
the box' value.  The total luminosity from the whole system including the
cylindrically expanding, long PCG outside the box will be discussed in
\S \ref{sec:lcurve}.

The overall evolutionary aspects of $R_{th}$ and $R_{r}$ for other
runs are similar to those for run 7 qualitatively,
and are different only quantitatively.  First, run 1,
for which no radiation is considered and $\gamma=5/3$ is adopted,
has significantly less $R_{th}$ than run 7.  This is mostly due to
the dependence of the shock conditions on $\gamma$~value: from
one-dimensional analysis, $u_{\gamma=4/3}/u_{\gamma=5/3} = 64/49$ where $u$
is the post-collision internal energy density.  However, the actual $R_{th}$
ratio between two simulations is quite larger than 64/49, probably because the
different physical environments in the collision region initially induced
by different $\gamma$~values continuously make the subsequent collisional
evolutions of the two simulations differ.

Runs 8 and 9 have $\Delta$'s different from that of run 7.
As shown in the previous section, the collision of run 9 is more of
bouncing while that of run 8 is more of merging, which provides more effective
thermalization environment.  This explains relatively higher $R_{th}$ value
of run 8.  However, interestingly, the $R_{r}$ values of these runs
and run 7 are nearly the same despite of considerably different structures
in the collision region.  This is because the less effective
collision forms a less dense PCG, which quickly radiates thermalized
energy due to a smaller optical depth.  The relatively high $R_{th}$ and
low $R_{r}$ values of run 10 can be attributed to the same reason.
On the other hand, both $R_{th}$ and $R_{r}$ of run 2 converge to
the values very close to those of run 7.  This implies that the stream height
is a relatively minor factor in determining $R_{th}$ and $R_{r}$.
Finally, the amounts of both thermalization and radiation decrease as the
z offset increases.

%%%%%%%%%%%%%%%%%%%%%%%%%%%%%%%%%%%%%%%%%%%%%%%%%%%%%%%%%%%%%%%%%%%%%%%%%%%%%%%%
\section{DISCUSSION}
\label{sec:discussion}

\subsection{Orbital Energy and Angular Momentum}

The collision changes the kinetic energy and angular momentum of the streams.
For the calculation of the latter, one needs the distance from the
collision region to the black hole, $r_{bh}$, and the angle between the
velocity vector of the stream 1 and the vector toward the black hole from
the collision region, $\theta_{bh}$.  We have adopted $r_{bh} = 6000 \rsun =
4.2 \times 10^{14} \, \rm{cm}$ from \markcite{K94}Kochanek (1994) and
$\theta_{bh} = 16\arcdeg$ from the requirement of the angular momentum
conservation.  These parameters depend on the mass of the black hole and
the collision impact parameter.

\begin{figure*}[p]
%Fig 10
\centerline{\epsfxsize=15cm\epsfbox{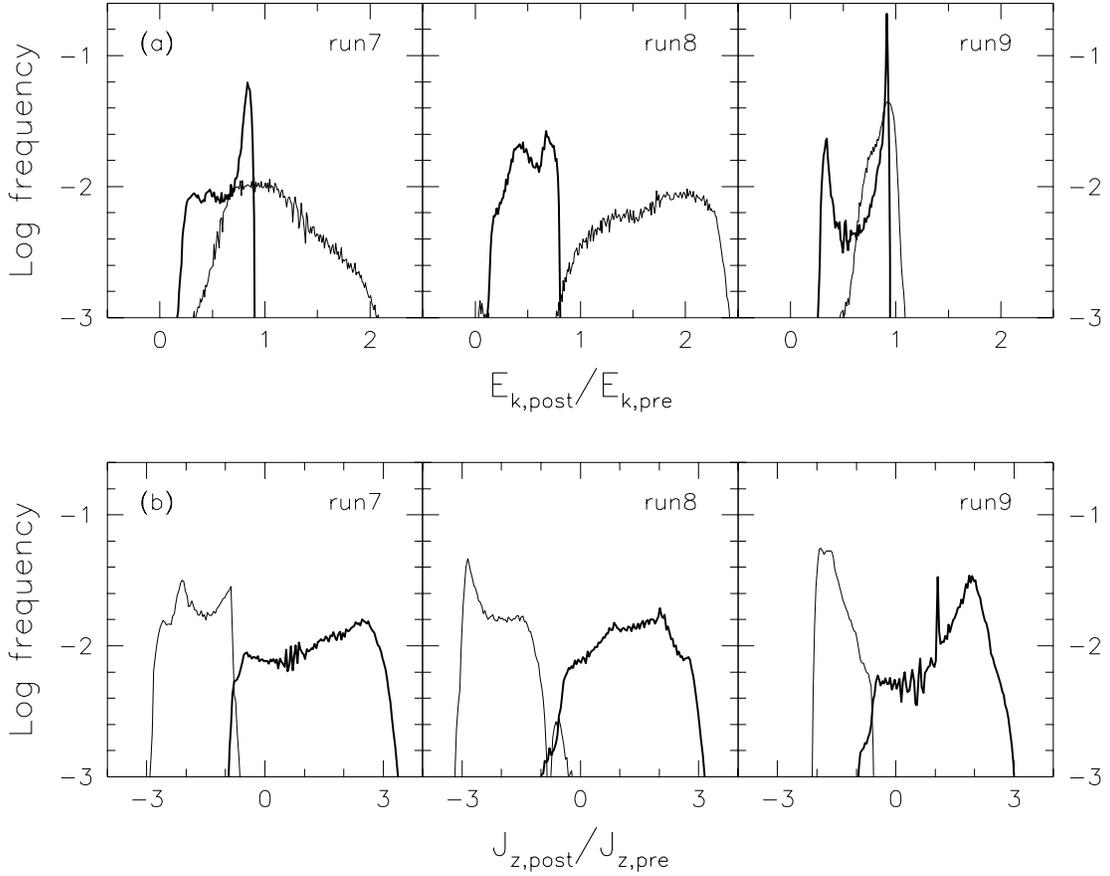}}
\caption
{\label{fig:EJhist}Normalized, density-weighted frequency distribution of
post-collision kinetic energy $E_{k,post}$ (a), and z-component post-collision
angular momentum $J_{z,post}$ (b).
Thick lines are for the gas with an angle between +y axis
and its velocity vector projected onto x-y plane, $\phi$, smaller than
$225 \arcdeg$ (PCG A), thin lines for the gas with $\phi$ larger than
$225 \arcdeg$ (PCG B).  $E_{k,post}$ and $J_{z,post}$ of PCG A (B) are
in units of pre-collision kinetic energy of stream 1 (2), $E_{k,pre}$, and
pre-collision angular momentum of stream 1 (2), $J_{z,pre}$, respectively.
The frequency distributions of PCGs A and B are normalized separately.}
\end{figure*}

The density-weighted frequency distribution functions (DFs) of post-collision
kinetic energy $f(E_{k,post})$ and those of z component of the angular momentum
$g(J_{z,post})$ for runs 7, 8, and 9 are shown in Figure \ref{fig:EJhist}.
Since the orbital evolution of the stream material is determined by $E_k$
and $J_z$ values in the post-collision expanding phase, only the cells
whose distances from the collision
center are within a certain range ($30 < r/l_{cell} < 40$) and whose velocity
vectors are heading away from the collision center were considered in the
calculation of the DFs.  Furthermore, since the expanding aspects of PCGs
A and B are considerably different, we have calculated two separate DFs for
the material with an angle between +y axis and velocity vector projected onto
the x-y plane, $\phi$, smaller than 225 $\arcdeg$ (PCG A) and for the material
with $\phi$ larger than $225 \arcdeg$ (PCG B).  The $f(E_{k,post})$ and
$g(J_{z,post})$ of PCG A (B) are normalized such that the pre-collision kinetic
energy $E_{k,pre}$ and z component angular momentum $J_{z,pre}$ values of
stream 1 (2) become unity, respectively.

Generally, $f(E_k)$ of PCG A are narrower and the peak is closer to
the unity (less affected by the collision) than those of PCG B.  As discussed
in \S \ref{sec:morphology}, this is because stream 1 has greater $\dot M/dA$.
The DF peak deviations from the unity of different runs are well explained
by the morphology of the collision regions.  While most material of
PCG A of run 8, which provides the strongest collision geometry, has
20-30\% less $E_k$ than stream 1, only few percent of stream 1 of run 9,
which has the weakest collision geometry, has experienced the $E_k$ reduction
of only about 10~\%.  In runs 7 and 8, PCG B has gained $E_k$
during the strong collision process.  On the other hand, in run 9, most
of PCG B has lost $E_k$ rather than gained.  This is  probably
because the bouncing phenomena with a small angle at the collision region
in run 9 does not effectively transfer the $E_k$ from stream 1 to
stream 2 and because the longer shock front between two streams has
preferentially thermalized stream 2 which has much smaller $\dot M/dA$ than
that of stream 1.

Unlike $f(E_{k,post})$, $g(J_{z,post})$ of runs 7, 8, and 9 shows similar
aspects with slightly different peak locations.  Streams initially have the
same (positive) sign for $J_z$, but the collision produces negative $J_z$,
resulting an increase of the total
absolute angular momentum.  In general, PCG has 2-3 times greater $|J_z|$
than the initial values $J_{z,i}$.  Hence the collision changes $J_z$ of
the stream much more easily than it changes $E_k$, which confirms
\markcite{K94}Kochanek's (1994) estimate.  This relatively large change in
$J_z$ is due to the rapid growth of the sine function (in $m r_{bh} v \sin
\theta_{bh}$) at small angle ($\theta_{bh}=16 \arcdeg$).
The average and the standard deviation of $J_{z,post}$ of PCG A in run7 are
1.73~$J_{z,i}$ and 0.87~$J_{z,i}$, respectively, where only the gas elements
whose sign of $J_{z,post}$ is the same as $J_{z,i}$ were considered.
Note that all the gas elements in our simulations are bound to the black
hole even after a part of them gains some kinetic energy through the
collision.  This is because the simulations here are for the very beginning
part of the debris stream, which is already well bound before the collision
due to 1) its proximity to the black hole during disruption and 2) hydrodynamic
effects that convert the orbital energy of the stream into internal energy
during the revolution around the black hole.

The exact details of the circularization process is still uncertain.
If the angular momentum is conserved during the circularization process,
the process will form the debris into a torus with a radius of
$j_{post}/GM_{bh}$, where $j_{post}$ is the angular momentum per mass of
the PCG.  Then the time scale for that process will depend on the difference
between the initial orbital (kinetic + potential) energy per mass of the
PCG and the energy per mass in a circular orbit of the same angular
momentum, $\Delta e$:
\begin{equation}
\label{delta_e}
        \Delta e = e_{post,i} + {1 \over 2} \left ( {GM_{bh} \over j_{post}}
			\right ) ^2
\end{equation}
where $e_{post,i}$ is the initial orbital energy per mass of the PCG.
The stream-stream collision lessens $\Delta e$ by
changing their pre-collision kinetic energy into thermal energy
(thus decreasing orbital energy) and increasing angular momentum.
We find that for runs 7, 8, and 9, the collision lessens $\Delta e$ by
a factor of 3 to 10, compared to the values which the stream would have
without the collision.  Thus if the angular momentum is conserved during
circularization process, the time scale for that process will be shortened
significantly by the collision, but the quantitative estimation of such
decrease in the time scale is still a difficult problem.

\subsection{Luminosity Evolution}
\label{sec:lcurve}

The luminosity from the gas is simply given by sum of the cooling rate,
equation (\ref{emit1}), over the whole volume. 
The luminosity from the gas within the simulation box $L_{box}(t)$ for
run 7 is plotted in Figure \ref{fig:lcurve}(a) with a solid line.  This curve
differs from $R_r$ of Figure \ref{fig:energy8b} in a sense that the latter is
time-cumulative energy ratio inside the box.  After a steep rise up to
$\sim 10^{41} \, {\rm erg \, sec}$ in $\sim 10^4$~sec, the luminosity
decreases to the steady state value of $\sim 10^{40} \, {\rm erg \, sec}$.
However, since the PCG continues to radiate beyond the simulation box, the
radiative cooling outside the box should also contribute to the total
luminosity.

To describe the behaviour of the gas outside the simulation box,
we use simple physical arguments.
The behaviour of the PCG outside the box will be
determined by the competition between the energy loss by radiation and that
by adiabatic expansion. Suppose adiabatic expansion is dominant over
radiative cooling. Since the outgoing PCG outside the box
is not bounded by incoming
streams or shocked gas, it may be described by the simple 
pressure driven expansion of a homogeneous gas cloud:
\begin{equation}
\label{dpdw}
	\left ( {P \over P_0} \right ) ^{-1/\gamma} \, dP
		= - \rho_0 \dot w \, d \dot w,
\end{equation}
where $P$ and $\rho$ are the pressure and the density inside the PCG,
$\dot w$ is the time derivative of the width of the PCG slice, $w$, and
subscript~$0$ is for the initial values.
Due to the one-dimensional nature of the PCG, we expect the 
expansion to be cylindrical:
\begin{equation}
\label{rho-w}
	{\rho \over \rho_0} = \left ( {w \over w_0} \right )^{-2}.
\end{equation}
Combining equations (\ref{dpdw}) and (\ref{rho-w}), one obtaines the
equation of motion for $w$:
\begin{equation}
\label{dwdt}
	{dw \over dt} = 2 \gamma P_0 \rho_0^{-1} w_0^{2\gamma-2} w^{1-2\gamma}.
\end{equation}
Here, the initial width of the PCG slice is defined by
\begin{equation}
\label{w0}
	w_0 \equiv (\pi \rho_0 v_{PCG} / \dot M_{PCG})^{-1/2},
\end{equation}
where $v_{PCG}$ and $\dot M_{PCG}$ are the stream velocity and the mass flux
of the PCG.

\begin{figure*}[p]
%Fig 11
\centerline{\epsfxsize=8.8cm\epsfbox{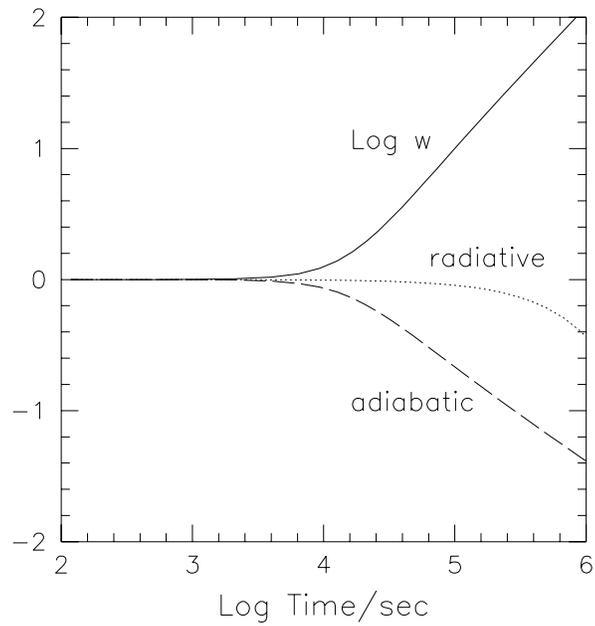}}
\caption
{\label{fig:wevol}Evolution of the width of the PCG slice $w(t)$ (solid).
The PCG slice is assumed to be a homogenous gas cloud that expands
adiabatically and cylindrically.  Initial conditions are obtained from
the results of run 7.  Also plotted are the internal energy evolution of
the PCG slice by pure adiabatic expansion, $\propto u(t)w(t)^2$ (dashed),
and that by pure radiative cooling, $\propto \exp(-t/t_d)$ (dotted).
All variables are plotted in logarithmic scale and normlized to their
initial values.}
\end{figure*}

\begin{figure*}[p]
%Fig 12
\centerline{\epsfxsize=15cm\epsfbox{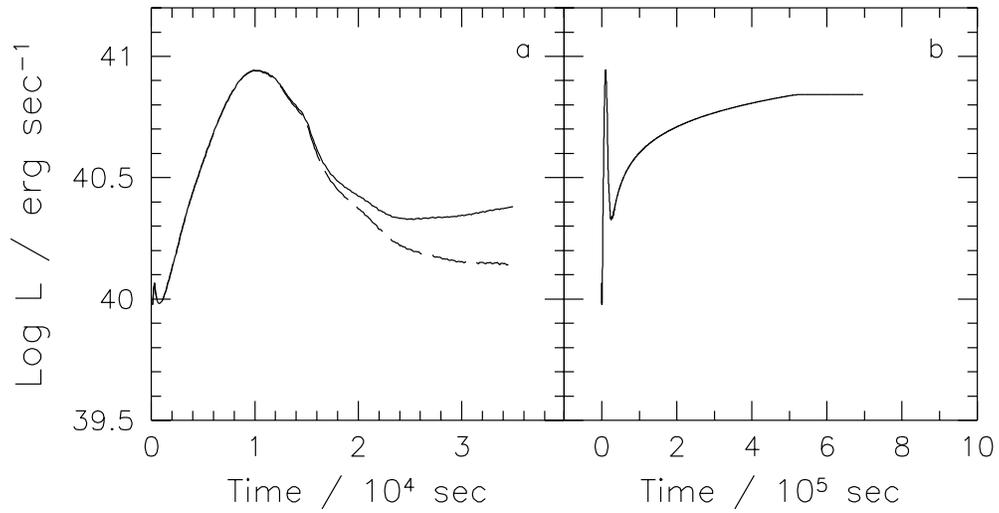}}
\caption
{\label{fig:lcurve}(a) Temporal evolution of the luminosity inside the
simulation box (solid line) and the total system luminosity including the
estimation for the PCG (dashes line) for run 7.  (b) Extrapolation of
total luminosity evolution with the physical quantities being the steady
state values at the last phase of the simulation.  A slice of the PCG
is assumed to adiabatically expand in two dimensions and to be luminous
only for $t_{dyn}$ after it leaves the collision region.}
\end{figure*}

Numerical integration of equation (\ref{dwdt}) with the initial values
from the results of run 7 is plotted in Figure \ref{fig:wevol} (solid line).
The width $w(t)$ grows slowly in the beginning, but soon attains the 
constant expansion velocity which is the consequence of equation (\ref{dpdw}).
Also plotted are the evolution of the internal energy in the slice
by pure adiabatic expansion $u(t)w(t)^2$ (dashed line) and by pure
radiative cooling $u_0 \exp(-t/t_d)$ (dotted line),
where $t_d \propto \tau_{es} w \propto nw^2$ is constant for cylindrical
expansion.  The former decreases much faster than the latter.
Therefore, the internal energy will be mainly
spent to the adiabatic expansion and the PCG may well be
described by the pressure driven adiabatic expansion.

It takes approximately $t_{dyn} \simeq 5 \times 10^5$~sec for the PCG to reach
the black hole from the collision region.  We found from the above $w(t)$
calculation that $\gamma_{eff}$ remains very close to $4/3$ at least for
$t_{dyn}$, and thus radiation energy will dominate over gas energy during
that period.  When the PCG passes near the black hole,
it could be significantly distorted by the tidal force and adiabatic
expansion may not be a good approximation because the tidal effect
could heat or cool the PCG if it passes very close to the hole.
However, according to the $w(t)$ calculation, the PCG by
then would have cooled down to $T \sim 3 \times 10^4$~K.
At this temperature
line opacities will dominate over the continuum ones and
the PCG will be very optically thick and radiates very little.
Since we already assumed that the pre-collision gas stream maintained near
this temperature does not radiate, 
we assume a slice of the PCG is luminous only 
for $t_{dyn}$ after it leaves the collision region
and ignore its contribution to the total luminosity afterwards.
%Also, the tidal effects may not be significant if the PCG
%has expanded enough when it passes near the hole.

The total luminosity, $L$, is the sum of the radiation from the gas
within the simulation box, $L_{box}$, and that from the gas
outside the box. For the emission from the PCG outside the box, we use
the results from the $w(t)$ calculation.  The emission rate per
unit length slice along the PCG, $l_{slice}$, is $\pi w^2 \epsilon$.
Since $\tau_{es} w$ is constant for cylindrical expansion, the emission
rate becomes
\begin{equation}
\label{lslice}
	l_{slice}(s) \propto u(\Delta t)w(\Delta t)^2,
\end{equation}
where $s$ is the distance along the PCG from the boundary of the simulation
box, $\Delta t$ is the time after the PCG leaves the boundary, and
$s=v_{PCG} \Delta t$.  Proportional constant is fixed to match the condition
at the simulation boundary, and the luminosity from the PCG outside the box
is now the sum of the slice emission along the whole PCG.  Thus the total
luminosity of the system is
\begin{equation}
\label{lcurve}
        L(t) \simeq L_{box}(t) + \int_0^{s_{max}} l_{slice}(s) \, ds.
\end{equation}

This total luminosity is plotted in Figure \ref{fig:lcurve}(a) with
a dashed line.  While the luminosity inside the simulation box (solid line)
converges to an asymptotic value, the total luminosity grows as the length
of the thermalized PCG increases.  We may extrapolate
the total luminosity evolution by assuming a steady state for the
gas inside the simulation box for $t>3 \times 10^4$~sec.
Figure \ref{fig:lcurve}(b) shows such extrapolation, where the total
luminosity reaches at $t \simeq t_{dyn}$ a steady state value of
$\sim 10^{41} \, {\rm erg \, sec^{-1}}$.  Since $t_{dyn}$ is about a half
of the collision phase, the durations of the exponential luminosity increase
and the steady state will be about the same.

The collision between the incoming stream and largely expanded PCG after its
passing the black hole for the second time will be highly ineffective.
This ineffective collision phase will last until the stream, slightly perturbed
by such an inefficient collision, passes the black hole and collides with its
tail at farther distance from black hole than the previous collision.
Since the velocity dispersion of the stream while passing the black hole
results in a larger cross section for more distant collision, subsequent
collisions will be less and less energetic.  The luminosity evolution
with a shape of Figure \ref{fig:lcurve}(b) will be repeated intermittently,
but with decreasing intensities.  After the debris is circularized,
the viscous accretion process will determine the evolution of the
luminosity from the system.

Emitted photons from the PCG will look Planckian because the gas is effectively
optically thick.  Since the temperature is less sensitive to the expansion of
the PCG ($T \propto n^{1/3}$), the characteristic temperature of the whole PCG
will not be far from that inside the simulation box (order of $10^5$~K).
However, it is certain that this characteristic temperature decreases as
the length of the PCG increases.

%%%%%%%%%%%%%%%%%%%%%%%%%%%%%%%%%%%%%%%%%%%%%%%%%%%%%%%%%%%%%%%%%%%%%%%%%%%%%%%%
\section{SUMMARY}
\label{sec:summary}

We have performed simulations of supersonic collisions between two gas
streams produced through the tidal disruption of a normal star by a super
massive black hole with a three-dimensional hydrodynamics code based on the
TVD scheme.  A simple treatment for radiative cooling valid in any optical
depth was included in the calculation.

Thermalized gas emerges out of the narrow, slab-like collision region
and expands out into a larger volume in two opposite directions, forming two
expanding cylinders several times thicker than the pre-collision
streams.  The angle of the shock slab, which determines the effectiveness
of the collision, depends primarily on the mass fluxes per unit area of the
colliding streams.
When there is no Lense-Thirring precession, approximately 10\% of the
initial kinetic energy of the streams is converted into thermal energy
during the collision depending on the stream parameters, and 3 to 4\% of
the thermal energy is immediately radiated away at the collision area
in form of radiation.  Off-centered collision which may be caused by
Lense-Thirring precession can somewhat reduce the amount of thermalization
and radiation, but still significantly alters the angular momentum vector
of the outgoing stream.

The collision alters the angular momentum of the stream more easily than it
changes the kinetic energy.  The angular momentum of the incoming stream is
increased by a factor of 2 to 3, and such increase, together with the decrease
in kinetic energy, reduces $\Delta e$, the difference between the
initial orbital energy per mass of the PCG and the energy per mass in a
circular orbit of the same angular momentum, by a factor of $\sim 10$.

The luminosity evolution of the system has three phases:  the initial
burst phase due to effective collision in the beginning, exponential
increase phase when the length of the thermalized, adiabatically expanding
post-collision gas increases, and steady state phase that lasts until the end
of the collision phase.  The radiation from the system is expected to be
close to Planckian with effective temperature of $\sim 10^5$~K.

\acknowledgements
We are grateful to Hyesung Kang, Hee-Won Lee, and Dongsu Ryu for their
helpful discussions.  The TVD code used in this study was kindly provided
by Dongsu Ryu and Hyesung Kang.  We also thank the anonymous referee for
comments which helped us improve the content of the discussion.  This work was
supported in part by the Cray Research and Development Grant in 1996 and
in part by Basic Science Research Institute Program to Pusan National
University, BSRI 97-2413.  M.-G. Park was supported by KOSEF Grant
971-0203-013-2.

%%%%%%%%%%%%%%%%%%%%%%%%%%%%%%%%%%%%%%%%%%%%%%%%%%%%%%%%%%%%%%%%%%%%%%%%%%%%%%%%
\begin{deluxetable}{ccccccc}
\tablecolumns{7}
\tablewidth{0pt}
\tablecaption{Stream Parameters
\label{table:stream}}
\tablehead{
\colhead{} &
\colhead{$\Delta_1$} &
\colhead{$\Delta_2$} &
\colhead{$H_1$} &
\colhead{$H_2$} &
\colhead{z offset} &
\colhead{} \\
\colhead{Run} &
\colhead{($l_{code}$)} &
\colhead{($l_{code}$)} &
\colhead{($l_{code}$)} &
\colhead{($l_{code}$)} &
\colhead{($l_{code}$)} &
\colhead{$\gamma$}
}
\startdata
run 7  & 5 & 3 & 5 & 3 & 0. & 4/3 \nl
run 8  &12 & 3 & 5 & 3 & 0. & 4/3 \nl
run 9  & 3 &12 & 5 & 3 & 0. & 4/3 \nl
run 2  & 5 & 3 & 5 & 5 & 0. & 4/3 \nl
run 10 &10 & 6 &10 & 6 & 0. & 4/3 \nl
run 11 & 5 & 3 & 5 & 3 & 1.6 & 4/3 \nl
run 12 & 5 & 3 & 5 & 3 & 3.2 & 4/3 \nl
run 1  & 5 & 3 & 5 & 3 & 0. & 5/3 \nl
\tablecomments{$l_{code}$ is the unit length used in the code and is equal
to ${5 \over 2} l_{cell} = 1.48 \times 10^{11}$~cm.}
\enddata
\end{deluxetable}

%%%%%%%%%%%%%%%%%%%%%%%%%%%%%%%%%%%%%%%%%%%%%%%%%%%%%%%%%%%%%%%%%%%%%%%%%%%%%%%%

%%%%%%%%%%%%%%%%%%%%%%%%%%%%%%%%%%%%%%%%%%%%%%%%%%%%%%%%%%%%%%%%%%%%%%%%%%%%%%%%
\clearpage

\clearpage

\clearpage

\clearpage

\clearpage

\clearpage

\clearpage

\clearpage

\end{document}